\documentclass[12pt]{article}
\usepackage[english]{babel}
\usepackage[square,numbers]{natbib}
\usepackage{url}
\usepackage[utf8x]{inputenc}
\usepackage{amsmath}
\usepackage{graphicx}
\graphicspath{{images/}}
\usepackage{parskip}
\usepackage{fancyhdr}
\usepackage{vmargin}
\setmarginsrb{3 cm}{2.5 cm}{3 cm}{2.5 cm}{1 cm}{1.5 cm}{1 cm}{1.5 cm}

\usepackage{amsmath}
\usepackage{amssymb}
\usepackage{graphicx}
\usepackage{qtree}
\usepackage{algpseudocode}

\title{Fragment Allocation Configuration in Distributed Database Systems}                                
\author{MR Abbasifard}                               
\date{10 July 2016}                                         

\makeatletter
\let\thetitle\@title
\let\theauthor\@author
\let\thedate\@date
\makeatother

\begin{document}


\begin{titlepage}
    \centering
    \vspace*{0.5 cm}
    \includegraphics[scale = 0.5]{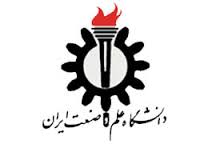}\\[1.0 cm]  
    \textsc{\Large Iran University of Science and Technology}\\[2.0 cm]  
    \textsc{\large MODB-201607DDB}\\[0.5 cm]               
    \rule{\linewidth}{0.2 mm} \\[0.4 cm]
    { \huge \bfseries \thetitle}\\
    \rule{\linewidth}{0.2 mm} \\[1.5 cm]
    
    \begin{minipage}{0.5\textwidth}
        \begin{flushleft} \large
            Mohammad Reza Abbasifard\\
            PhD Candidate\\
            \end{flushleft}
            \end{minipage}~
            \begin{minipage}{0.5\textwidth}
            \begin{flushright} \large
            Omid Isfahani Alamdari \\
            Research Associate\\
        \end{flushright}
        
    \end{minipage}\\[2 cm]

\end{titlepage}


\begin{abstract}
In distributed database (DDB) management systems, fragment allocation is one of the most important components that can directly affect the performance of DDB. In this research work, we will show that declarative programming languages, e.g. logic programming languages, can be used to represent different data fragment allocation techniques. Results indicate that, using declarative programming language significantly simplifies the representation of fragment allocation algorithm, thus opens door for any further developments and optimizations. The under consideration case study also show that our approach can be extended to be used in different areas of distributed systems.

\end{abstract}

\pagebreak

\tableofcontents
\pagebreak


\section{Introduction}

\begin{figure}
  \centering
  \includegraphics[scale=.3]{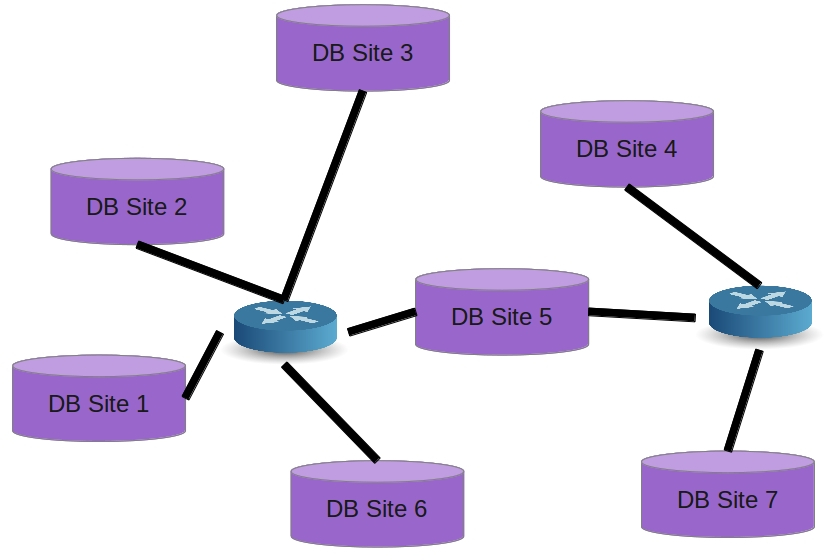}
  \caption{An example of a DDS.}
  \label{fig:ddb_01}
\end{figure}

Developments in distributed algorithms, network technologies, and database theory in the past few decades led to advances in distributed database systems (DDS). A DDS is a collection of database nodes connected by a communication network, in which each node is a database system in its own right, but the nodes have agreed to work together, so that a user at any node can access data anywhere in the network exactly as if the data were all stored at the user's own node (See Figure~\ref{fig:ddb_01}). 

The primary concern of fragmentation in a DDS is to show how data should be divided and distributed among nodes in the underlying database. Fragmentation problem in a DDS is how to divide the data while allocation issue means how those fragments should be distributed over different DDS nodes. The data allocation problem, is NP-complete, and thus requires fast heuristics to generate efficient solutions \cite{DBLP:journals/tods/MeghiniT91}. Furthermore, the optimal allocation of database objects highly depends on the query execution strategy employed by a distributed database system, and the given query execution strategy usually assumes an allocation of the fragments.

A major cost in executing queries in a distributed database system is the data transfer cost incurred in transferring relations (fragments) accessed by a query from different nodes to the node where the query is initiated. The objective of a data allocation algorithm is to determine an assignment of fragments at different nodes so as to minimize the total data transfer cost incurred in executing a set of queries. This is equivalent to minimizing the average query execution time, which is of primary importance in a wide class of distributed conventional as well as multimedia database systems. 

An optimal, but not practical, solution for fragment allocation in DDS has been appeared in \cite{Morgan:1977:OPD:359581.359591}. There are also a few fragment allocation algorithms \cite{Basseda2009,BassedaTasharofiNNA,DBLP:conf/aiccsa/BassedaR09,DBLP:conf/icsnc/BayatiGRB06,Ahmad2002,Corcoran:1994:GAF:326619.326738} that are proven to be practical and show a reasonable performance. Several surveys of those algorithms are provided by \cite{DBLP:conf/pdis/MuthurajCVN93,DBLP:conf/sigmod/NavatheR89,DBLP:journals/tods/NavatheCWD84,Apers:1988:DAD:44498.45063,Brunstrom:1995:EED:221270.221652,BassedaTasharofiTechReport}. Since all of these fragment allocation algorithms are expressed and implemented by imperative programming languages, they are usually difficult to understand and configured.

In this paper, using declarative rule based languages, we propose a novel technique that can be used to represent fragment allocation algorithms. In our technique, we consider fragment allocation strategy as a rule-based policy, implemented in a logic programming framework. The declarative representation of fragment allocation algorithms results in two major benefits: (1) since declarative representation of algorithms are much simpler than imperative ones, these algorithms can be changed and improved simpler when they are represented by rule-based languages; (2) the reasoning components of these algorithms can be relied on logic programming frameworks, and thus we will have simpler implementation of fragment allocation components in DDS. This technique also can be used to improve existing DDS fragment allocation simulators \cite{BassedaTasharofiSimulatorManual}.

\begin{figure}
  \centering
  \includegraphics[scale=.4]{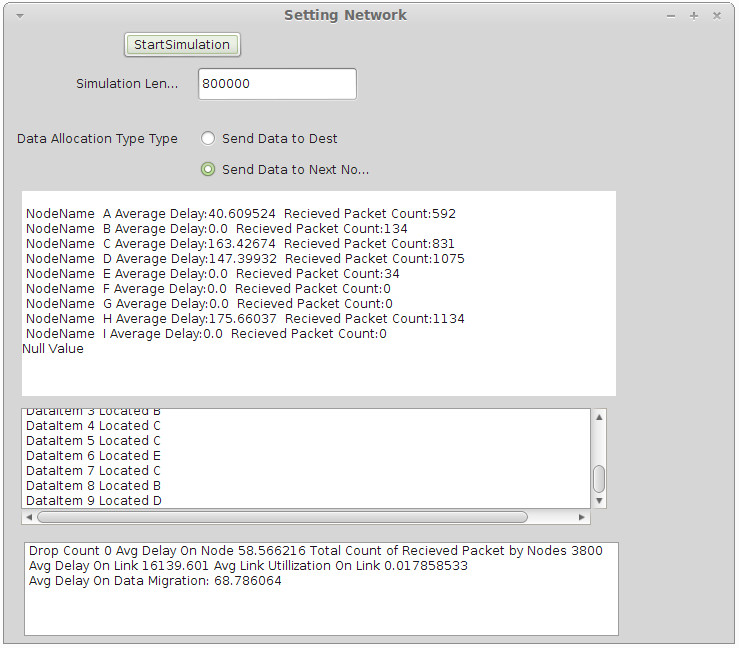}
  \caption{GUI of DDB simulator in \cite{BassedaTasharofiSimulatorManual}.}
  \label{fig:BassTashSim}
\end{figure}

The rest of this paper is structured as follows: Section~\ref{sec:graph-modeling} shows how we can model a DDS as a graph. in Section~\ref{sec:problem-desc}, we will briefly review some of major parameters of fragment allocation problem. Section~\ref{sec:methodlogy} is about our representation technique and Section~\ref{sec:implementation} briefly explains the implementation of our prototype model. Finally Section~\ref{sec:conclusion} draws our conclusion. 

\section{Modeling a DDS as a Graph} \label{sec:graph-modeling}

\begin{figure}
  \centering
  \includegraphics[scale=.3]{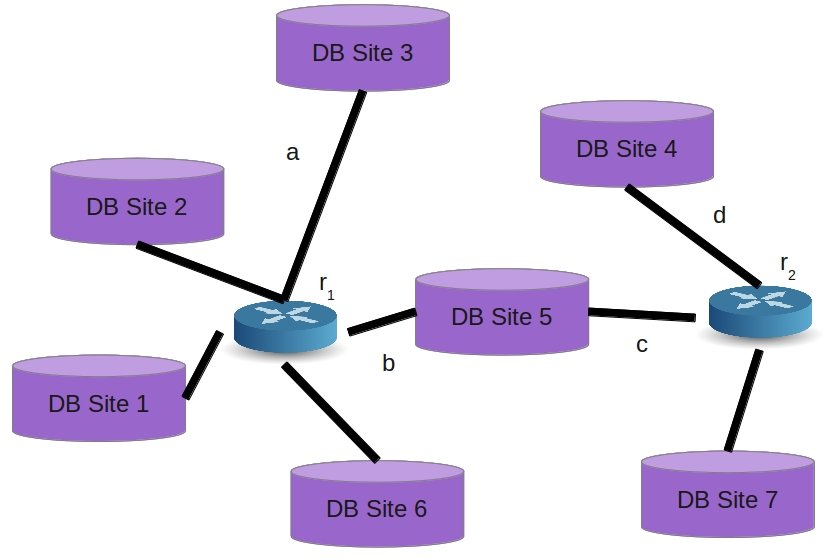}
  \caption{An example of a DDS --- routers, edges, and sites.}
  \label{fig:ddb_02}
\end{figure}

In this section, using an example, we will show how a DDS can be represented and model as a graph. The following modeling technique first has been introduced by \cite{BassedaTasharofiSimulatorManual}. We will have a brief overview of this technique to make this report self-contained and the details of this modeling is not in the scope of this report. Consider the DDS shown in Figure~\ref{fig:ddb_01}. Let some of nodes, routers, and edges of that DDS be identified as shown in Figure~\ref{fig:ddb_02}. For each $ i $, an element of this system (i.e. edge, site, router), let $ \delta(i) $ denote the delay of $ i $ and $ \omega(i) $ be its assigned bandwidth. In order to make our models as simple as possible, without loss of generality, we assume that:

      \begin{equation}
      \forall i \in Edges \cup Routers, \omega(i) = +\infty
      \end{equation}

\begin{figure}
  \centering
  \includegraphics[scale=.3]{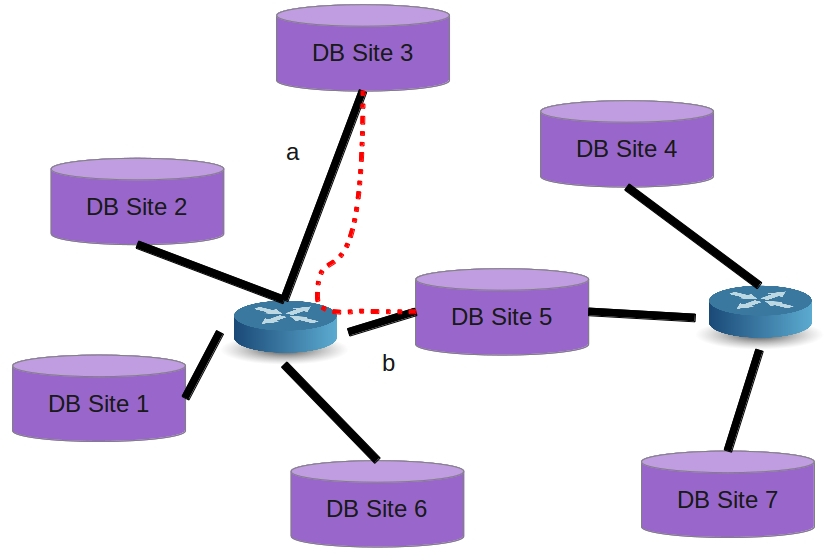}
  \caption{Drawing a path between \emph{DB Site 5} and \emph{DB Site 3}.}
  \label{fig:ddb_03}
\end{figure}

Clearly, for every pair of sites $ i $ and $ j $ that are connected through a set of routers, one can assume a connecting edge and compute the corresponding delay and bandwidth. For instance, as shown in Figure~\ref{fig:ddb_03}, one can draw a path between \emph{DB Site 5} and \emph{DB Site 3} and assume an edge between those sites. Let $ x_{a+b} $ denote the hypothetical edge between those sites. Then, one can show that $ \omega(x_{a+b}) = min\{\omega(a),\omega(b)\} $ and $ \delta(x_{a+b}) = \delta(a) + \delta(r_1)  + \delta(b)$. 

\begin{figure}
  \centering
  \includegraphics[scale=.3]{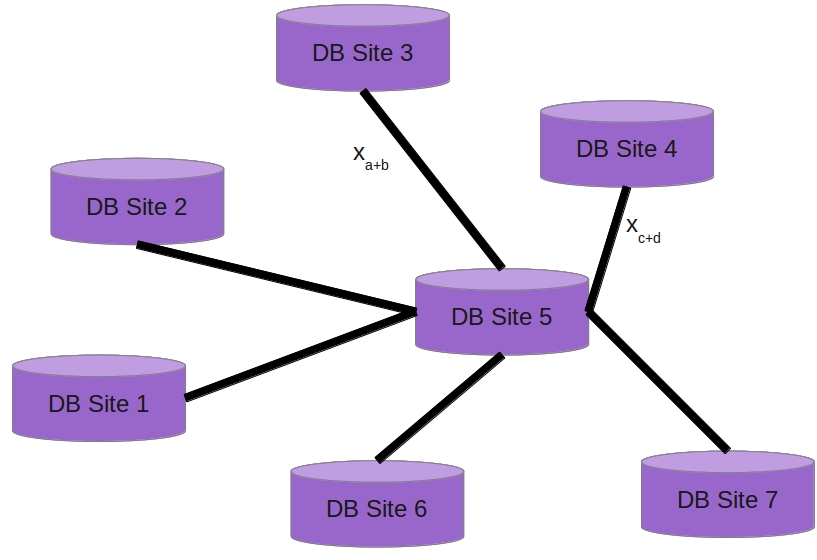}
  \caption{The graph model of the DDS shown in Figure~\ref{fig:ddb_01}.}
  \label{fig:ddb_04}
\end{figure}

Removing routers from a DDS, one can draw a simpler model to study different fragment allocation algorithm. For instance, Figure~\ref{fig:ddb_04} shows the graph model of the DDS shown in Figure~\ref{fig:ddb_01}.  

\section{Fragment Allocation Problem} \label{sec:problem-desc}

Fragment and data allocation algorithms are categorized into two major groups: static and dynamic. In static fragment allocation algorithms, data allocation has been completed prior to the design of a database depending on some static data access patterns and/or static query patterns. However, dynamic fragment allocation algorithms can change the data fragment allocation automatically during the deployment of the database. In a dynamic environment where these probabilities change over time, the static allocation solution would degrade the database performance.

Depending on the complexity of a data allocation algorithm, it may take the following parameters as inputs: 
\begin{enumerate}
\item The fragment dependency graphs.
\item Unit data transfer costs between nodes.
\item The allocation limit on the number of fragments that can be allocated at a node.
\item The query execution frequencies from the nodes.
\end{enumerate}

\begin{figure}
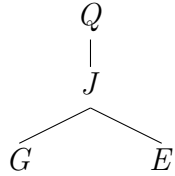
%
\bigskip
\Tree [.\textit{Q} [.\textit{J} \textit{G} !\qsetw{3cm} \textit{E} ] !\qsetw{3cm} ]
\medskip
\caption{A sample fragment allocation graph.}
\label{fig:graph}
\end{figure}

The fragment dependency graph models the dependencies between the fragments and the amount of data transfer incurred to execute a query. A fragment dependency graph (as shown in figure 1) is a rooted directed acyclic graph with the root as the query execution site (Node $ Q $ in Figure \ref{fig:graph}) and all other nodes as fragment nodes (Node $ G $, etc., in Figure \ref{fig:graph}) at potential nodes accessed by a query.

Assume that $ r_{ij} $ indicates the frequency of requirements by node $ i $ for fragment $ j $, each fragment $ i $ is characterized by its size, $ n_i $ and $ t_{ij} $ indicates the cost for node $ i $ to access a fragment located on node $ j $. Clearly, $ t_{ij} $ is a function of the following parameters: 

\begin{itemize}
\item The average size of data fragments: $ s_j $.
\item The bandwidth of network link between $ i $ and $ j $: $ w_{ij} $.
\item The delay of network link between $ i $ and $ j $: $ d_{ij} $.
\item Other types of costs on network link between $ i $ and $ j $, e.g. communication expenses: $ o_{ij} $.
\end{itemize}

\begin{figure}
  \centering
  \includegraphics[scale=.5]{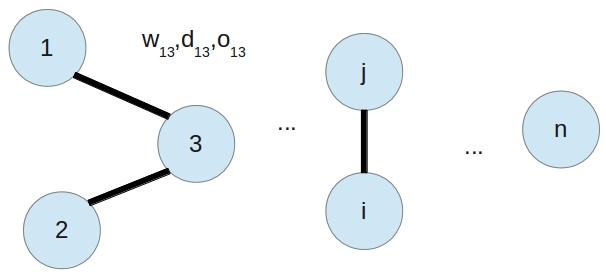}
  \caption{A sample network parameters.}
  \label{fig:network}
\end{figure}

Therefore, users of the distributed database systems must be able to define $ t_{ij} $ for a fragment allocation algorithm based on the above mention parameters. Moreover, the frequency of the execution of each type $ k $ of the queries executed by node $ i $ on data item $ j $, $ f_{ijk} $, is another important factor for the fragment allocation algorithm. Note that, different types of database queries have different transfer costs. For instance, \emph{select} ($ se $) queries (specially those require joins on tables) may require large data transfers while \emph{update} ($ up $) and \emph{delete} ($ de $) queries do not require large data transfers. In fact, an efficient fragment allocation algorithm results in minimization of execution cost, which is shown in~(\ref{eq:execution_cost}).

\begin{equation}
\sum\limits_{k \in \{se,up,de \}}\sum\limits_{i=1}^{m}\sum\limits_{j=1}^{n} f_{ijk}
\label{eq:execution_cost}
\end{equation}
 
The distributed database allocation problem is to find the optimal placement of the fragments at the nodes. That is, we wish to find the placement, $ P= \{p_1,p_2,p_3,\ldots,p_j,\ldots,p_n \} $ (where $ p_j = i $ indicates fragment $ j $ is located at node $ i $) for the $ n $ fragments so that the capacity of any node is not exceeded, that is shown in~(\ref{eq:capacity_limit}).
\begin{equation}
\sum\limits_{i=1}^{m} r_{ij} n_j \leq c_{ij}
\label{eq:capacity_limit}
\end{equation}

Moreover, the total transmission cost, shown in~(\ref{eq:transmission_cost}), should be minimized \cite{Corcoran:1994:GAF:326619.326738}.

\begin{equation}
\sum\limits_{i=1}^{m}\sum\limits_{j=1}^{n} r_{ij} t_{ij}
\label{eq:transmission_cost}
\end{equation} 

By restricting the use of the requirements matrix and having zero transmission cost, the distributed database allocation problem can be transformed to the bin packing problem, which is known to be NP-complete.

\section{Methodology} \label{sec:methodlogy}

In this paper, our goal is to develop a flexible and dynamic fragment allocation algorithm. Clearly, such algorithm must be considered as a distributed algorithms. Otherwise, adding a coordinator node can drastically decrease the flexibility of such algorithm. At the first glance, developing such distributed algorithm may look difficult as distributed logic programming and rule based frameworks are required for such algorithm. But, fortunately, this problem is not as difficult as what it looks. Because synchronizing the fragment allocation and its parameters, each node can act independently while we make sure the result of our executions for different nodes are same. Then, we just need to represent our fragment allocation algorithm using a rule based language and make sure the rules of each node and facts are properly synchronized. 

\begin{figure}

\begin{verbatim}
delay(1,3,5).
...
reverse_bandwidth(1,3,0.5).
...
other(1,3,5).
\end{verbatim}
\caption{Representation of network as a set of facts.}
\label{fig:net_rep}
\end{figure}

In order to develop a fragment allocation algorithm in a rule-based language, first we need to represent above mentioned parameters as sets of facts. Then, we need to develop our algorithm in terms of rules---similar to representation of policies using rule based languages. Obviously, the set of rules defining the fragmentation algorithm should be synchronized in each node as well.

The over all representation of network parameters in a rule based language is simple and natural. We can use simple sets of facts to represent $ s_j $, $ w_{ij} $, $ d_{ij} $, and $ o_{ij} $. For instance, Figure~\ref{fig:net_rep} shows that the delay between node $ 1 $ and $ 3 $ is $ 5 $ milliseconds, the reverse of the bandwidth is $ 05 $ $ 1/ $mega-bytes, and the cost of communication for each mega-byte is $ 5 $ dollars. Then, $ t_{ij} $ can be computed as shown by~(\ref{eq:t_computation}), where $ \gamma_{ij} $ represents the user defined factors. This computation will be translated to a rule in our algorithm. Figure~\ref{fig:t_rule} shows a sample translation of such computation.

\begin{equation}
t_{ij} = \gamma_{ij} \times s_j \times  w_{ij} \times d_{ij} \times o_{ij}
\label{eq:t_computation}
\end{equation}

\begin{figure}

\begin{verbatim}
transfer_cost(I,J,T) :- user_defined_parameter(I,J,U),
                        size(J,S),
                        reverse_bandwidth(I,J,W),
                        delay(I,J,D),
                        other(I,J,O),
                        T is U*S*W*D*O.
\end{verbatim}
\caption{Representation of the computation of $ t_{ij} $ in our algorithm.}
\label{fig:t_rule}
\end{figure}

Similarly, the execution statistics, $ f_{ijk} $ can also be generated as a set of fact by the execution engine of DDS. The pre-defined parameter to show the execution cost of query type $ k $ on node $ i $ for the fragment $ j $, $ e_{ijk} $, is also defined as a fact by users. Therefore, for the simplest fragment allocation policy, where fragments are moved if the execution cost is larger than fragment relocation cost. In such algorithm, the trigger for moving the data item $ j $ from $ i_1 $ to $ i_2 $, $ move_{i_{1}i_{2}j} $, can be computed through the following rule:

\begin{align}
\label{eq:move_rule}
move_{i_{1}i_{2}j} \longleftarrow &  \sum\limits_{k \in \{se,up,de\}} f_{i_{1}jk} \leq  r_{i_{1}j} t_{i_{1}j} ~\wedge~ \\ \nonumber  &\sum\limits_{k \in \{se,up,de\}} f_{i_{2}jk}  >  r_{i_{2}j} t_{i_{2}j} 
\end{align}

Accordingly, this trigger runs two major events: physically moving the data item $ j $ from $ i_1 $ to $ i_2 $ and updating fragment allocation information in all of the nodes. Using rules of type~(\ref{fig:t_rule})~and~(\ref{eq:move_rule}), the inference engine needs to respond to the query~(\ref{eq:move_query}), where $ X $, $ Y $, and $ Z $ are variables bound by inference engine. The result of such query will be used to activate triggers.

\begin{equation}
?-~ move_{X,Y,Z}.
\label{eq:move_query}
\end{equation}

Simply, one can use prolog $ assert $ and $ retract $ instructions in synchronization unit to update fragment allocation information. Based on this executions, the main procedure of fragment allocation component can be developed as shown in Figure~\ref{alg:frag_main}.

\begin{figure}
\begin{algorithmic}[1]
\Function {FRAGMENT\_ALLOCATION}{}
\While {true}
\State Run synchronization unit
\State Update execution statistics
\If {Any facts updated}
\State Re-run the inference engine and query the $ move_{X,Y,Z} $ triggers.
\If {There exists any trigger whose source is me}
\State Run the fragment transfer unit
\EndIf
\Else
\State Wait for synchronization period
\EndIf
\EndWhile

\EndFunction
\end{algorithmic}
\caption{The main procedure in fragment allocation component.}
\label{alg:frag_main}
\end{figure}

As mentioned before, rule based representation of fragment allocation algorithm makes those algorithms simple and easy to understand. For instance, let $ a_{i_{1}i_{2}} $ be a fact representing that there is a direct link between $ i_1 $ and $ i_2 $. Therefore, NNA \cite{BassedaTasharofiNNA} fragment allocation algorithm can be simply represented as

\begin{align}
move_{i_{1}i_{2}j} \longleftarrow &  \sum\limits_{k \in \{se,up,de\}} f_{i_{1}jk} \leq  r_{i_{1}j} t_{i_{1}j} ~\wedge~ \\\nonumber  &\sum\limits_{k \in \{se,up,de\}} f_{i_{2}jk}  >  r_{i_{2}j} t_{i_{2}j} ~\wedge~ \\ \nonumber  & a_{i_{1}i_{2}}
\label{eq:move_NNA_rule}
\end{align}

Similarly, FNA \cite{DBLP:conf/aiccsa/BassedaR09}\cite{Basseda2009} and BGBR \cite{DBLP:conf/icsnc/BayatiGRB06} parameters can be imported to our algorithms. Complicated reasoning for FNA also needs supporting Fuzzy logic resolutions and libraries by resolution frameworks.

\section{Implementation} \label{sec:implementation}

\begin{figure}
  \centering
  \includegraphics[scale=.20]{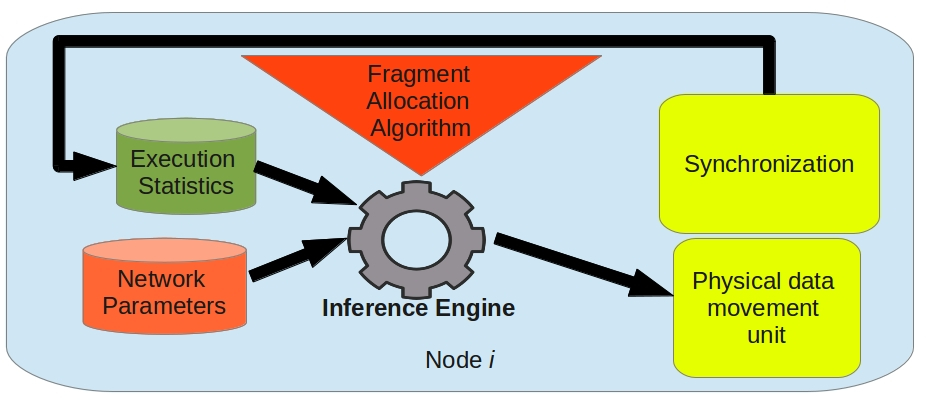}
  \caption{Design of a single node in a DDS.}
  \label{fig:node}
\end{figure}

As mentioned in the previous section, in our approach, each node is considered as an independent system, synchronized with other nodes on fragment allocation mechanisms. Figure~\ref{fig:node} shows the design of a node in our DDS. We are still working on the implementation of this project. The inference engine in our system will be XSB Prolog \cite{xsb-swift-2011}. The implementation will be evaluated using the parameters introduced in \cite{Basseda2009,BassedaTasharofiNNA}.

\emph{Synchronization} is one of the most important components of our system. Synchronization is repeated in a period of time. The frequency of synchronization also depends on the speed of the execution of fragment allocation algorithm by inference engine. Apparently, each node must wait until receive the synchronization information from the rest of the nodes before each execution of the fragment allocation algorithm.

\section{Conclusion} \label{sec:conclusion}
In this paper, we discussed a novel method for representing fragment allocation algorithms in a rule based system. Our results show that such representation makes a fragment allocation algorithm. The simplicity of the resulted algorithm can help one to extend existing algorithms and improve their performances. Moreover, the simplicity of the resulted algorithms eases configuring fragment allocation component in DDS.

We are planning to investigate using defeasible reasoning and argumentation theory \cite{DBLP:conf/iclp/WanGKFL09}\cite{DBLP:conf/ruleml/BassedaGKGC15} to extend our developments. Another promising direction for this research is to investigate other rule based system, e.g. Answer Set Programming \cite{DBLP:conf/iclp/GelfondL88}\cite{DBLP:journals/aicom/GebserKKOSS11} , and possibly get more speedups.

\newpage
\bibliographystyle{plain}
\bibliography{RuleBasedDDBFragmentAllogcation_extended}

\end{document}